\documentclass[aps,prd,showpacs]{revtex4}
\usepackage{graphicx}
\def\mnras{Mon.\ Not.\ R.\ Astron.\ Soc.\ }
\begin{document}
\title
{Amplification of azimuthal modes with 
odd wave numbers during dynamical bar-mode growth
in rotating stars}
%
\author{Yasufumi Kojima}
\email{kojima@theo.phys.sci.hiroshima-u.ac.jp}
%
\affiliation
{Department of Physics, Hiroshima University, Higashi-Hiroshima
  739-8526, Japan}
%
\author{Motoyuki Saijo}
\email{saijo@rikkyo.ac.jp}
%
\affiliation
{Department of Physics, Rikkyo University,
Toshima, Tokyo 171-8501, Japan}
%
\date{8 September 2008}
%
\begin{abstract}
  Nonlinear growth of the bar-mode deformation is 
studied for a differentially rotating star with supercritical 
rotational energy. In particular, the growth mechanism of 
some azimuthal modes with odd wave numbers is examined
by comparing a simplified mathematical model with a realistic simulation.
Mode coupling to even modes,
i.e., the bar mode and higher harmonics, significantly 
enhances the amplitudes of odd modes, unless
they are exactly zero initially.
Therefore, other modes which are not axially symmetric 
cannot be neglected 
at late times in the growth of the unstable bar-mode 
even when starting from an almost axially symmetric state. 
\end{abstract}
%
\maketitle
%
\section{Introduction}
%
There are many three-dimensional simulations 
that have been carried out for dynamically unstable 
modes in rotating stars for decades
\citep{TDM,DGTB,WT,HCS,SHC,HC,TIPD,NCT,LL,Liu02,
SK08,SBS00,SSBS01,BPMR07,CNLB01,SBS03,OT06,CQF07,OTM07}. 
The calculations describe the onset of the instability 
for almost axially symmetric states
and its evolution in the nonlinear regime.
The critical rotational parameter determining the bar-mode 
instability, i.e., the ratio of the rotational 
$T$ and gravitational binding energies $W$,
has been shown to depend weakly on the
rotation law and the equation of state, 
both in Newtonian
\citep{TDM,DGTB,WT,HCS,SHC,HC,TIPD,NCT,LL,Liu02,SK08} 
and relativistic gravity\citep{SBS00,SSBS01,BPMR07}. 
In addition, recent discoveries from the numerical simulations suggest
that dynamical bar instabilities can happen at significantly small
values of rotation, 
which are associated with high degree of differential
rotation \citep{TH90,PDD,SKE,CNLB01,SBS03,OOTB05,OT06,CQF07}.  
Some of the numerical findings can be triggered by the corotation
resonance for the so-called low 
$T/W$ dynamical instability \citep{WAJ05,SY06}, which is completely different
in nature from the high $T/W$ one.
We here consider the dynamical instability with
large rotational parameter, high $T/W$ instability only.

   The fate of the unstable bar mode is an interesting problem.
The timescale for the persistence of the bar shape
is very important for the detection of gravitational waves.
The problem can only be solved using time-dependent numerical codes 
which require high resolution and accuracy 
to follow the long-term evolution without being
overwhelmed by numerical errors.
This problem has been independently attacked by various 
groups using different computational codes.
Typically, a small non-axisymmetric initial perturbation is added
to a rotating axisymmetric model which is in unstable equilibrium
and the subsequent evolution is calculated numerically.
The results show that the bar structure is destroyed 
in a dynamical timescale, i.e., 
within a few multiples of the rotational period,
after the amplitude attains nonlinear 
saturation (see e.g.,\citep{SK08,BPMR07}). 

  In these simulations,
some azimuthal modes with odd wave numbers appear in the later growth
of the bar shape (which has azimuthal number $m=2$)
\footnote{
In this paper, a 'mode' is not used in 
a rigorous meaning, but rather means a Fourier component
in the azimuthal direction.
}.
It is, therefore, interesting to 
study the physical mechanism of the growth of odd modes.
Are there unstable modes with odd number (e.g., $m=1$ or $3$)
in addition to the unstable bar mode?
Are the amplitudes of the odd modes enhanced
by mode coupling?
Three-dimensional simulation of hydrodynamics, even in Newtonian
gravity, is time-consuming for a wide range of initial data and
parameter sets. So the system is too complicated for extracting the
physical mechanism.
In order to understand the growth of odd modes, we investigate
the evolution of a simplified model. The model's description of
mode coupling, unstable growth and decay mimics the realistic system 
very well. 
Moreover, the number and growth rates of the unstable modes
are easily controlled. The model, therefore, deepens our 
understanding of the nonlinear behavior of 
unstable bar-mode growth in rotating stars.
The physical mechanism is confirmed 
by comparing the model problem with a more realistic calculation of a
dynamically unstable star simulated using 
three-dimensional hydrodynamics in Newtonian gravity.

  The organization of this paper is as follows. 
Recent numerical results for 
three-dimensional hydrodynamics in Newtonian gravity
are summarized in Sec.~\ref{sec:3Dsimulation}. 
A simplified model is presented in Sec.~\ref{sec:Toymodel}.
We investigate how odd modes grow within 
our mathematical model
and check that the conclusion is consistent with 
three-dimensional numerical results. Finally a discussion of our 
results is given in Sec.~\ref{sec:Discussion}.

\section{Three-dimensional Newtonian Hydrodynamics}
\label{sec:3Dsimulation}
\subsection{Methods}
The set of equations for three-dimensional hydrodynamics 
in Newtonian  gravity
is the continuity, Euler, energy and Poisson equations:
\begin{equation}
\partial_{t}\rho +\partial_{i} (\rho v^{i}) = 0,
\label{eqn:continuit}
\end{equation}
\begin{equation}
\partial_{t}(\rho v_{i})
+ \partial_{j} (\rho v_{i} v^{j})
=
- \partial_{i} (P + P_{\rm vis})
- \rho \partial_{i} \Phi
\label{eqn:Euler}
\end{equation}
\begin{equation} 
\partial_{t} (\rho \varepsilon)^{1/\Gamma}
+ \partial_{j}((\rho \varepsilon)^{1/\Gamma} v^{j})
= 
- \frac{1}{\Gamma}  (\rho \varepsilon)^{-(\Gamma-1)/\Gamma}
 P_{\rm vis} \partial_{i} v^{i},
\label{eqn:energy}
\end{equation}
\begin{equation}
\triangle \Phi = 4 \pi G \rho .
\label{eqn:Possion}
\end{equation}
Here we assume a $\Gamma$-law equation of state ($\Gamma =2$),
  in which thermal pressure $P$ is given by the density 
  $\rho$ and specific internal energy $\varepsilon$ as
\begin{equation}
P = ( \Gamma - 1 ) \rho \varepsilon .
\label{eqn:GammaLaw}
\end{equation}
Pressure $P_{\rm vis}$ in eqs.(\ref{eqn:Euler}) and (\ref{eqn:energy}) 
is the artificial viscosity pressure introduced
to deal numerically with shocks.
 The form (\ref{eqn:energy}) for the energy equation 
 is derived by eliminating  pressure term, and is convenient 
 for the numerical calculation.
The equations are numerically solved in Cartesian 
coordinates $(x,y,z)$, 
assuming planar symmetry across the equator.   
  We use PCG(preconditioned conjugate gradient) 
  method e.g, \cite{PCG} to solve the elliptic
  equation (\ref{eqn:Possion}) with parallel processors.
The 3D hydrodynamical simulation code in Newtonian
gravity has been developed, parallelized and tested in the context of
dynamical instabilities in Refs.~\cite{SBS03,SY06,SK08}.

As initial data, we construct differentially rotating equilibrium
models with the so-called $j$-constant rotation law 
with $d=1$ ($d$ is a parameter which represents the degree of
differential rotation) given by
\begin{equation}
\Omega = \frac{j_{0}}{R_{\rm eq}^{2} + x^{2}+y^{2}}.
\label{eqn:omega}
\end{equation}
Here $\Omega$ is the angular velocity around $z$-axis, 
$j_{0}$ is the constant parameter
with units of specific angular momentum, and
$R_{\rm eq}$ is the  stellar radius on equatorial plane.
For the construction of the density distribution
at equilibrium,
we also assume a polytropic equation of state with $n=1$ 
\begin{equation}
P = \kappa \rho^{2},
\end{equation}
where $\kappa$ is a constant.

  In the numerical calculations,
we monitor the azimuthal Fourier components
$m=1$, $2$, $3$, $4$ which are defined by
%
\begin{eqnarray}
D &=& \left< e^{i m \varphi} \right>_{m=1} 
\nonumber \\
&=&
  \frac{1}{M} \int \rho 
  \frac{x + i y}{\sqrt{x^{2}+y^{2}}} d^3 x
\label{eqn:dipole}
,\\
Q &=& \left< e^{i m \varphi} \right>_{m=2} 
\nonumber \\
&=&
  \frac{1}{M} \int \rho 
  \frac{(x^{2}-y^{2}) + i (2 x y)}{x^{2}+y^{2}} d^3 x
\label{eqn:quadrupole}
,\\
O &=& \left< e^{i m \varphi} \right>_{m=3} 
\nonumber \\
&=&
  \frac{1}{M} \int \rho 
  \frac{x (x^2 - 3 y^2) + i y (3 x^2 - y^2)}{(x^{2}+y^{2})^{3/2}} d^3 x
,\\
M_{4} &=& \left< e^{i m \varphi} \right>_{m=4} 
\nonumber \\
&=&
  \frac{1}{M} \int \rho 
  \frac{(x^4 - 6 x^2 y^2 + y^4) 
    + i (4 x y (x^2 - y^2))}{(x^{2}+y^{2})^2} d^3 x,
\end{eqnarray}
where $M$ is the total rest mass and an angular bracket 
denotes the density weighted average.

\begin{table}[htbp]
\begin{center}
\leavevmode
\caption{
Four different rotating equilibrium stars in Newtonian gravity of
$\Gamma = 2$.
}
\begin{tabular}{c c c c}
\hline
\hline
Model & $R_{\rm p} / R_{\rm eq}\footnotemark[1]$ & $T/W$ & 
stability of the bar mode
\\
\hline
  I & $0.225$ & $0.281$ & unstable
\\
 II & $0.250$ & $0.277$ & unstable
\\
III & $0.275$ & $0.268$ & unstable
\\
 IV & $0.300$ & $0.256$ & stable
\\
\hline
\hline
\end{tabular}
\label{tab:equilibrium}
\footnotetext[1]{$R_{\rm p}$: Polar radius; $R_{\rm eq}$: Equatorial radius}
\end{center}
\end{table}

\subsection{Results for Nonlinear Growth of Non-axially Symmetric Modes}
  The growth of the bar-type instability depends on
the rotational parameter $T/W$, 
the ratio of the rotational to gravitational binding energies.
The growth rate increases
with $T/W$ when it is larger than a critical value of
$T/W \approx 0.26 \sim 0.27$.
Our numerical study is limited to the instability
with such large rotational parameter. 
We start from hydrostatic equilibrium models for
gravity, centrifugal force and pressure gradient.
In our numerical experiments we found that initial models
with very large values of $T/W$ are not suitable 
for examining the unstable growth, since 
the growing mode  too rapidly destroys the bar shape.
Models with marginal parameter values are better and
we will discuss four models ($T/W =0.256 -  0.281$)
summarized in Table 1. 
%
To excite any dynamically unstable mode, we disturb the initial
equilibrium density $\rho_{\rm eq}$ with a non-axisymmetric perturbation
given by 
\begin{equation}
\rho = \rho_{\rm eq}
\left[ 1 + \sum _{m=2,3,4}   \delta^{(m)} Y_m (x,y)
\right],
\label{eqn:DPerturb}
\end{equation}
where $\delta^{(m)} $ is a small constant, and 
$Y_m $ is a polynomial of order $m$ given by
\begin{eqnarray}
Y_2 &=& \frac{1}{R_{\rm eq}^{2}}
\left[  x^{2} + 2 x y - y^{2} \right],
\\ 
Y_3  &=& \frac{1}{R_{\rm eq}^{3}} 
\left[ x( x^{2} -3y^{2}) +y(3 x^{2} -y^{2} ) \right],
\\
Y_4 &=&  \frac{1}{R_{\rm eq}^{4}}
\left[ x^4 - 6 x^2 y^2 + y^4 + 4 x y (x^2 - y^2)\right] .
\end{eqnarray}
The amplitude of $\delta^{(m)} $ does not coincide with
the diagnostics in the azimuthal Fourier components 
at the initial state, rather the
relation is given by e.g, 
$|Q| = |\delta^{(2)}| ( \int \rho_{\rm eq} (x^2+y^2) d^3 x ) 
/(\sqrt{2}M R_{\rm eq}^{2}) $.
In order to exclude behaviors originating from 
numerical errors,  
we check several conditions (the center of mass, 
conservation of linear momentum and conservation of 
angular momentum) throughout the evolution.
All of them are well conserved within several percent.
We also terminate our code once the relative error in the rest mass
exceeds $ 10^{-4}$, because the matter spread out from the
computational grid.
The detailed study for the numerical checks
and the accuracy of the numerical codes are
given in the previous paper\citep{SK08}.
Typically, the integration is terminated
at late times of the simulation for unstable cases,
where turbulent-like behavior is seen.
%

\begin{figure}
\centering
\includegraphics[keepaspectratio=true,width=14cm]{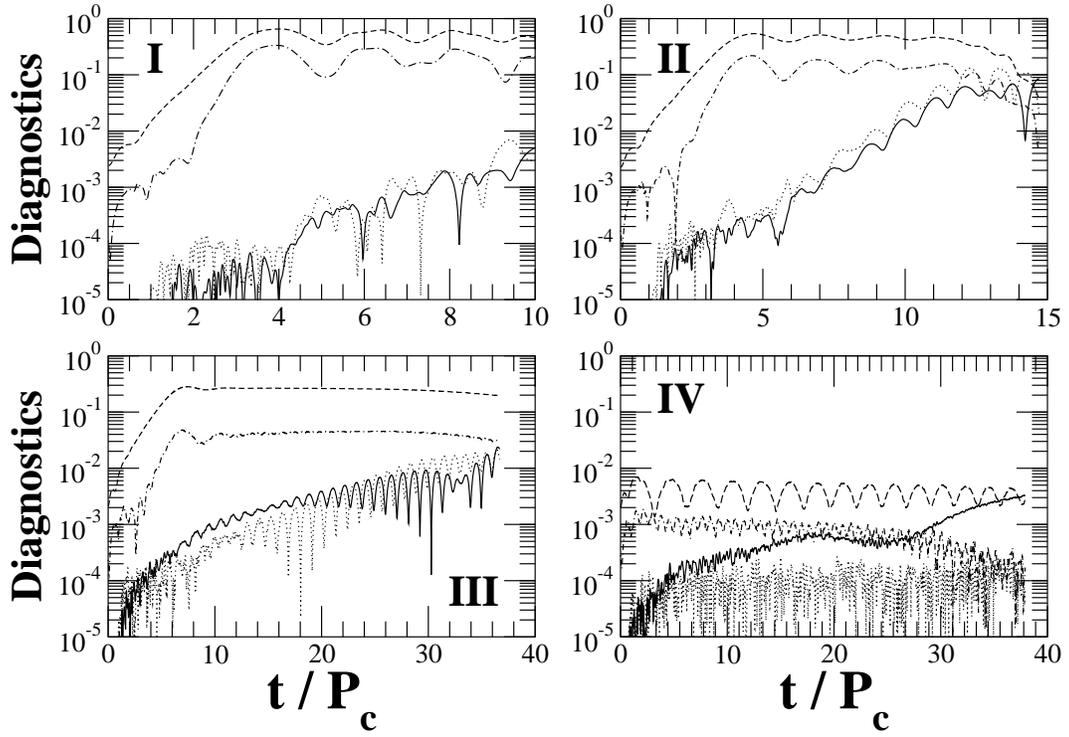}
\caption{
Diagnostics as a function of
$t/P_{\rm c}$ for four rotating stars I-IV where  
$P_{\rm c}$ is the central rotation period at $t=0$. 
Solid, dashed, dotted, and dash-dotted lines
denote the diagnostics for $|D|(m=1)$, $|Q|(m=2)$, 
$|O|(m=3)$ and $|M_4|(m=4)$, respectively. 
Models  I-III are unstable, whereas 
model  IV is  stable.}
\label{fig:diagnostic1}
\end{figure}

\begin{figure}
\centering
\includegraphics[keepaspectratio=true,width=14cm]{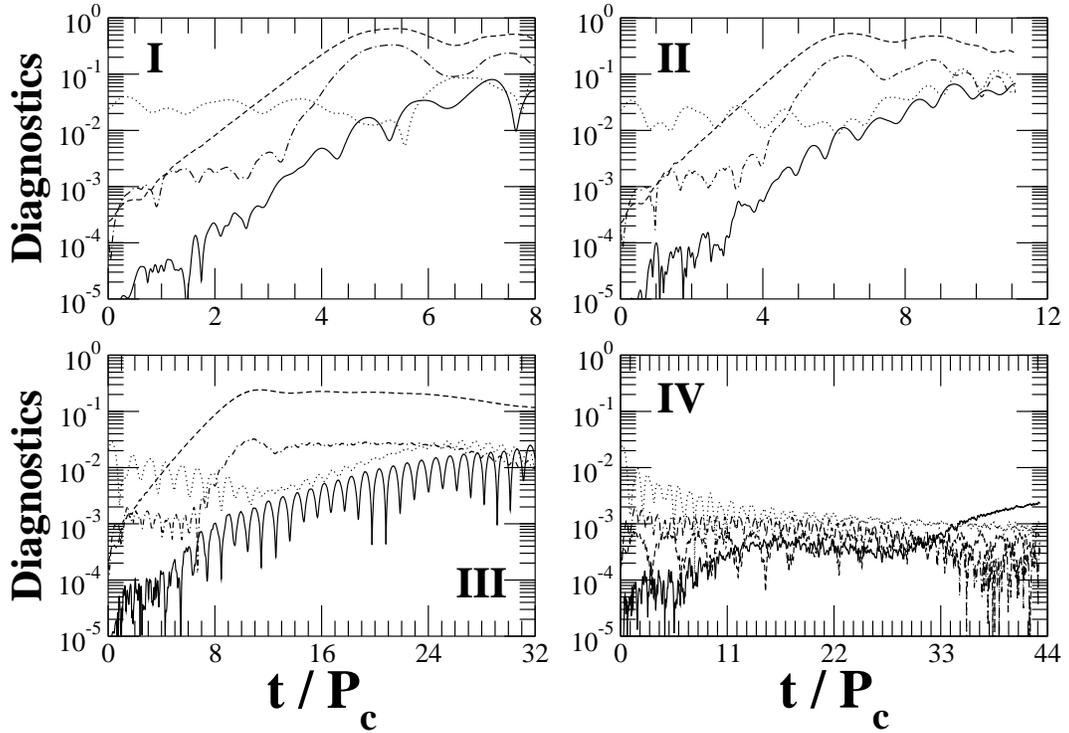}
\caption{
The same as Fig.1, but initial perturbations are
$\delta^{(2)} =\delta^{(3)}= 10^{-3}$,
$\delta^{(4)} =0$.}
\label{fig:diagnostic2}
\end{figure}

The evolution of the bar instability
is shown in Figure~\ref{fig:diagnostic1}.
The initial amplitudes of the perturbation
are  $\delta^{(2)} = \delta^{(4)} = 10^{-2}, \delta^{(3)}=0 $.
It is easily seen that models I-III are unstable,
whereas model IV is stable. 
The exponential growth of the $m=2$ mode
can be seen for models I-III, whereas 
an oscillation of $m=2$ with finite amplitude
$\sim 5 \times 10^{-3}$  can be seen in model IV.
The $m=4$ mode also grows exponentially 
because of the second harmonic of $m=2$
in the unstable models (I-III).

  The evolution of the bar instability for different initial
perturbation is shown in Figure~\ref{fig:diagnostic2}.
The rotational parameter is the same as that of 
Figure~\ref{fig:diagnostic1}, but
the initial amplitude is reduced to 
$\delta^{(2)} = 10^{-3} $, and that of odd mode is added
$\delta^{(3)}=10^{-3} $, $\delta^{(4)} = 0$.
Even for small initial amplitude of $m=2$, unstable
growth can be seen.
The overall features of the diagnostics for each
component are the same as those of Figure~\ref{fig:diagnostic1}.
The unstable bar mode grows in the models (I-III),
but  the saturation time, in which the amplitude of 
$m=2$ mode is  $ \sim {\cal O} (1)$,  is delayed due to
initially small amplitude.
In the stable system, model IV,
the amplitudes of all diagnostics become 
same order $ \sim {\cal O} (10^{-3})$.

 The interesting result is that the odd modes 
grow for late times in the unstable models,
irrespective of the initial perturbations.
The finite differencing scheme used in the numerical code
always generates  small fluctuations in all $m$ modes.
This ``noise'' level is expected to be less than $10 ^{-4}$,
or $10 ^{-3}$ at most by a rough estimate.
That is,
the relative, numerical error in the global quantities is
estimated as $1/N^2 \approx 10^{-4} $-$ 10^{-3}$, 
since second order scheme with typical grid number 
$400 \times 400 \times 100$ is adopted. 
%
We also cover the equilibrium star with $120$ diameter points in the
equatorial plane.
The amplitudes of odd modes 
exceed this level in the unstable models (I-III) 
in Figure~\ref{fig:diagnostic1}.
The amplitudes of $m=3$ in Figure~\ref{fig:diagnostic2}
exceed it from initial conditions, 
but they turn to growth after the saturation of
unstable mode.

\begin{figure}
\centering
 \includegraphics[keepaspectratio=true,width=16cm]{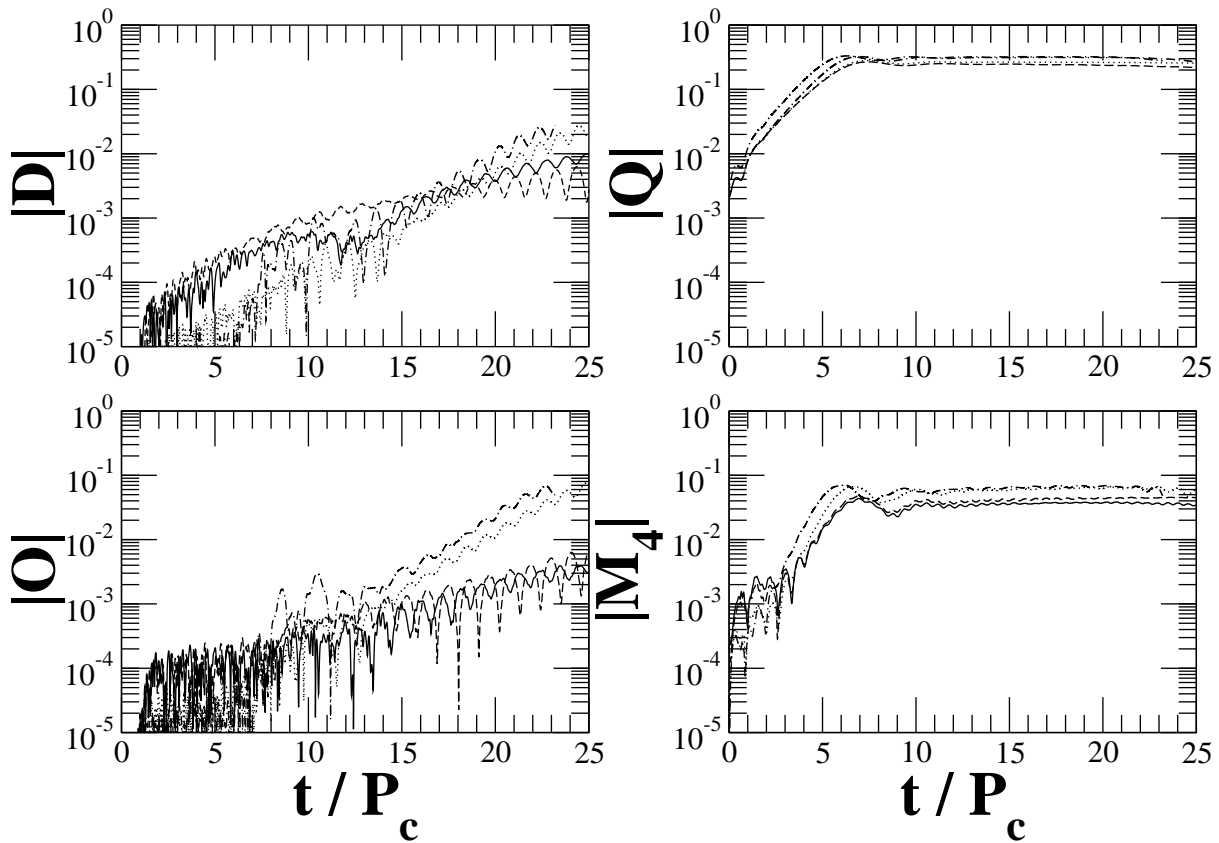}
\caption{
Each four diagnostics as a function of $t/P_{\rm c}$ with four
  different types of spatial computational resolution.  The
  computational grid 
  size is set as (a) ($320 \times 320 \times 80$) (solid line), (b)
 ($400 \times 400 \times 100$) (dashed line), (c) ($400 \times 400
 \times 100$) (dotted line), and (d) ($500 \times 500 \times 125$)
 (dash-dotted line), covering the  grid points of the equatorial
 diameter of the equilibrium star as (a) $96$, (b) $120$, (c) $240$,
 and (d) $300$.
}
\label{fig:resolution}
\end{figure}

In order to show that the result is not a numerical artifact with
  no physics, we have investigated four different types of spatial
  computational resolution.  We set the computational grid size as (a)
  ($320 \times 320 \times 80$), (b) ($400 \times 400 \times 100$), (c)
  ($400 \times 400 \times 100$), and (d) ($500 \times 500 \times
  125$), covering the equatorial diameter of the equilibrium star with
  (a) $96$, (b) $120$, (c) $240$, and (d) $300$ grid points.  Note
  that case (a) is 48.8\% ($= 100 \times (1 - 0.8^{3})$) reduction of
  the grid points of case (b), and that cases (c) and (d) have twice
  and $2.5$ resolution in the matter regime as 
  that of case (b) respectively, keeping the same grid size (the
  radius of the outer boundary is set closer in cases (c) and (d) than
  in case (b)).  We have found that the odd azimuthal modes grow
  exponentially at the late time in all four cases as shown in
  Figure~\ref{fig:resolution}.  The initial condition and equilibrium
  model are the same as those in  Figure~\ref{fig:diagnostic1}-III for
  cases (a), (b), and (c), but we put $1.5$ times large density
  perturbation for case (d).  However there are some quantitative
  differences among three resolutions, cases (a) and (b) have
  milder steeps in the exponential growth of odd modes than cases(c)
  and (d).  This might come from the diffusive effect of the large
  meshes 
  of the computational grid.  Hereafter, we will not discuss the
  precise growing feature, but rather discuss the mechanism of the odd
  azimuthal modes in the dynamical bar-instability.  In this case, our
  choice of the computational resolution (case(b)) can explain the
  growth of odd azimuthal modes
  for the above purpose.  The growth of the odd modes can be seen in both
  relativistic and present Newtonian results, and should be worthy of
  investigation.  In order to study it by further numerical works,
  different approaches or much sophisticated treatments would be
  necessary.  We here pay attention to a possible growth mechanism of
  the odd modes.
%

\section{One-dimensional Flow Coupled to a Scalar Field}
\label{sec:Toymodel}
\subsection{Mathematical Model} 
It is clear that the most important nonlinearity in 
eqs.(\ref{eqn:continuit})-(\ref{eqn:Possion})
is the advection term, $(v \nabla ) v $,
when written in non-conservative form.
Taking into account the nonlinearity and instability
caused by an external force,
we introduce a simplified model to
examine the nonlinear evolution of the unstable modes. 
Our model is Burgers' equation for 
a flow velocity $ u(t,x) $
coupled to a scalar field $\phi(t,x)$:
\begin{equation}
 \partial_{t}u+u\partial_{x}u=\nu \partial_{x} ^2 u
+\lambda \phi ,
\label{eqn:Burgers}
\end{equation}
\begin{equation}
 \partial_{x} ^2 \phi+2\phi =-u+1 ,
\label{eqn:Burgersforce}
\end{equation}
where $\nu$ is a diffusion constant.
It is well known that eq.(\ref{eqn:Burgers}) 
with $ \lambda =0 $ represents a shock model 
due to the nonlinear advection term $u\partial_{x} u$.
We assume $\nu >0 $ for stability.
The additional term $ \lambda \phi$ is
a force  mimicking gravity and  may cause
global instability, as shown below.
The scalar field $ \phi$ satisfies a second-order
partially differential equation (\ref{eqn:Burgersforce})
which is like the Poisson equation with a source $u$.
The additional term $2\phi $,  which may come from 
geometrical factors, such as curvature, in a realistic system, 
is introduced to adjust the $m=2$ mode to  the most unstable one.
See the next subsection.
We assume that the system 
of eqs.(\ref{eqn:Burgers}) and (\ref{eqn:Burgersforce})
is in non-dimensional form and that
the spatial range is limited to $0 \leq x \leq 2\pi$.
Periodic conditions are imposed on the
functions $u$ and $\phi$ at $ x=0, 2\pi$.
Therefore,  $ x$ corresponds to azimuthal angle $ \varphi$
in the realistic system.
It is easily checked that 
$1-u=\phi =0$ is an exact solution of eqs.
(\ref{eqn:Burgers}) and (\ref{eqn:Burgersforce}).
This solution corresponds to an axisymmetric solution 
for eqs.(\ref{eqn:continuit})-(\ref{eqn:Possion}).
We regard $u=1$ and $\phi=0$  as the background state 
and consider linear stability and nonlinear growth 
from this uniform state.

  We solve eqs.
(\ref{eqn:Burgers}) and (\ref{eqn:Burgersforce})
using Fourier series expansion for $0\leq x\leq2\pi$:
\begin{equation}
 u=1+ \sum_{m=1} 
a_{m}(t)\cos(mx)+b_{m}(t)\sin(mx).
\label{eqn:FourierForm}
\end{equation}
From eq.(\ref{eqn:Burgersforce}) we have
\begin{equation}
\phi =\sum_{m=1}
\frac{1}{m^{2}-2} \left( a_{m}(t)\cos(mx)+b_{m}(t)\sin(mx) \right).
\end{equation}

The number and growth rates of the unstable modes
are easily controlled  by changing the initial data and the parameters
$\nu$ and $\lambda$.
Our model is one-dimensional and is, therefore, easily
solved for a wide range of parameters.
We will show that this model's description of
mode coupling, unstable growth and decay mimics the realistic system of
eqs.(\ref{eqn:continuit})-(\ref{eqn:Possion}),
very well. 
 
%

\subsection{Linear Perturbation}
Assuming that  $|a_{m}|, |b_{m}| \ll 1$, 
we linearize eq.(\ref{eqn:Burgers}) giving
\begin{equation}
\frac{d}{dt}(a_{m}\pm ib_{m})\mp im (a_{m}\pm ib_{m})=\Gamma_{m} 
(a_{m}\pm ib_{m}),
\end{equation}
where
\begin{equation}
\Gamma _{m} \equiv -\nu m^{2}+\frac{\lambda}{m^{2}-2} .
\label{eqn:Gammadef}
\end{equation}
The solution can be written as
$ a_{m}\pm ib_{m} \propto \exp(\pm im t + \Gamma_{m} t) $, and 
the stability of mode $m$ is therefore
determined by the sign of $\Gamma_{m}$. That is, 
$\Gamma_{m}$ represents
the growth rate (for $\Gamma_{m}>0$) 
or the decay rate (for  $\Gamma_{m}<0$).
It is easily seen that the diffusion term with $\nu>0$
is stabilizing, while  the term $\lambda\phi$ with
$\lambda>0$ is de-stabilizing.
The growth rate depends on the magnitudes of $\lambda$ and $\nu$, 
whereas the number of unstable modes depends only on the ratio
of two constants $\lambda /\nu $.
The $m$ mode becomes unstable if $ \lambda/ \nu > m^2(m^2-2)$.
This model shows that  as $\lambda$ increases,
the short wavelength modes, i.e., those with large $m$, become unstable.
For example, all modes are stable for  $ \lambda / \nu  < 8$.
For $8 < \lambda /\nu < 63$ the only unstable mode is $m=2$.
For $ 63 < \lambda /\nu < 224$ the unstable modes are $m=2$ and $m=3$.

  However, the $m=1$ mode is always stable. 
This choice of the model may correspond to some constraints
in a realistic system, e.g., no motion of the center of mass.
Furthermore, the most unstable mode is adjusted to $m=2$
in this model. 
The correspondence to the realistic hydrodynamical system 
is clear with respect to wave number $m$, 
but we can not at present demonstrate that the system of 
eqs.(\ref{eqn:Burgers})-(\ref{eqn:Burgersforce})
is approximately derived from a realistic system.

\subsection{Numerical Calculation for Nonlinear Growth}
In this section  we use the Fourier series expansion (\ref{eqn:FourierForm}),
but do not assume that $a_{m}(t),b_{m}(t)$ are small.
From eq.(\ref{eqn:Burgers}) we have a coupled system of
ordinary  differential equations for $a_{m}(t)$ and $b_{m}(t)$
$(m =1,2, \cdots)$. 
In the numerical calculations,   
the range of $m$ is truncated to $  m_{\rm max}=40$.
The parameter $\lambda $ is set to
$\lambda =1$, so that the growth timescale is the dynamical one.
The viscous timescale associated with $\nu$ 
should be less than the dynamical one. 
The value $\nu$ is quite small in realistic situations,
but is not so small in the model calculation.
We used $\nu = 0.01 \sim 0.15$ to save simulation time 
and control the number of unstable modes.
For this choice of $\nu$, the viscous timescale 
$\sim \nu^{-1} m^{-2}$,  given by eq.(\ref{eqn:Gammadef}),
is larger than the dynamical one for small $m$ modes,
but not for large $m$ modes, e.g, $ m > \nu^{-1/2} $. 
The decay rate $\Gamma _m$ is, therefore, modified
so as to suppress the very rapid decay of large $m$ modes. 
The maximum decay rate is set to $ |\Gamma _m |=1$.

 It is well known that some finite difference schemes
for integrating the inviscid Burgers' equation, 
i.e., $ \nu=\lambda=0$ 
in eq.(\ref{eqn:Burgers}), become unstable at the shock\cite{H88}:
overstable oscillations with high frequency 
are generated.  
Our numerical scheme is not a finite difference scheme 
but we tested it for the case $ \nu=\lambda=0$ and
found the scheme to be stable.
One drawback of the numerical method is
the Gibbs phenomenon: an overshoot at the shock front.
The oscillation cannot be removed by increasing 
the number $m_{\rm max}$.
However, this peculiarity always appears at a discontinuity
even for a static problem.
For example, the discontinuity of a sawtooth form is 
poorly expressed by a Fourier sum\cite{AF01}.
Furthermore, note that the Gibbs phenomenon 
occurs with other eigenfunction expansions.

\begin{figure}
\centering
\includegraphics[keepaspectratio=true,width=16cm]{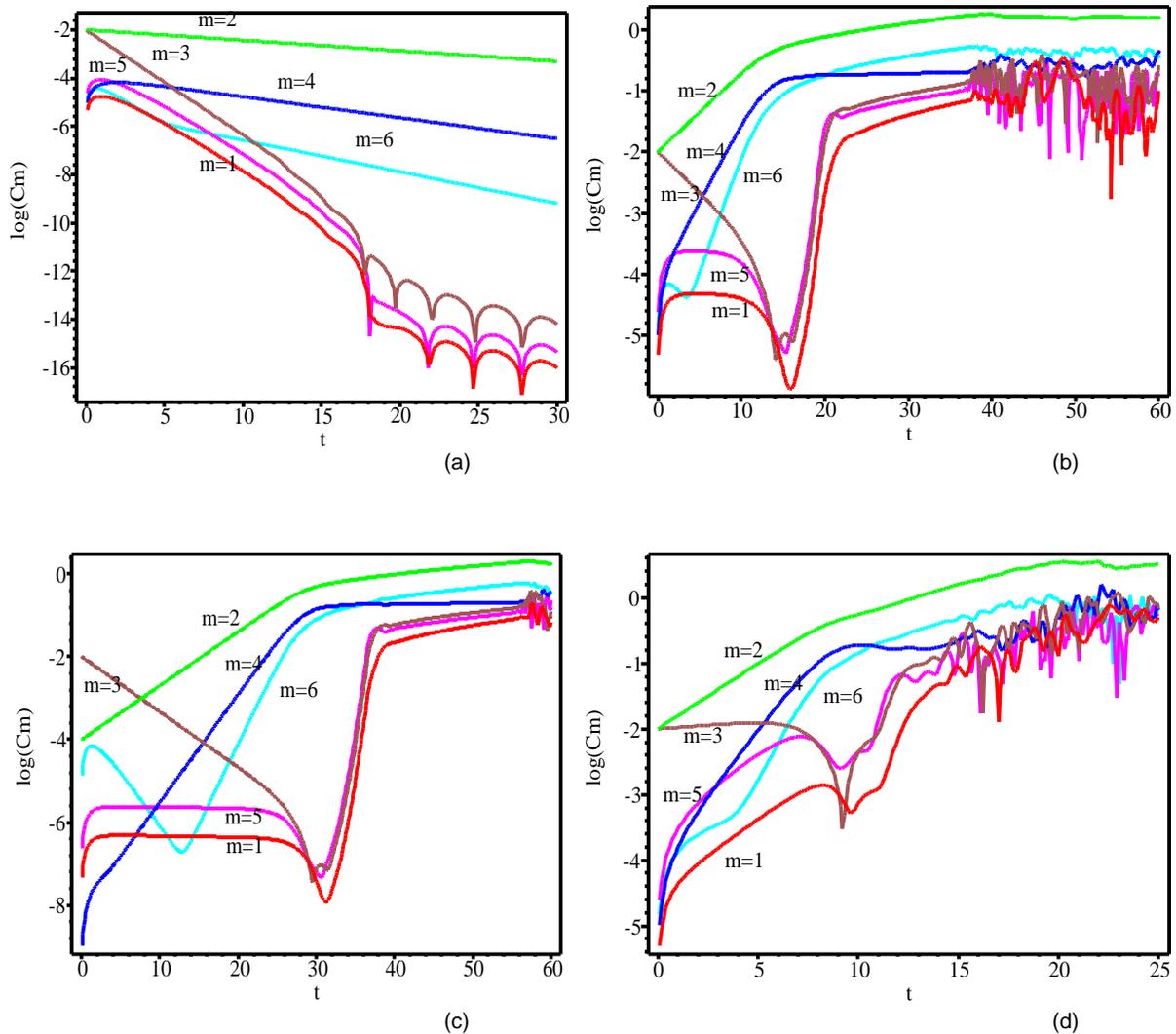}
 \caption{
The time-evolution of the amplitudes of
the lowest six Fourier modes $(m=1,\cdots 6)$.
Panel (a) is result for $\nu=0.15$,
(b) and (c) for $\nu=0.05$, and (d) for $\nu=0.01$.
The initial amplitudes are $a_2=a_3=10^{-2}$
for (a), (b) and (d), while 
$a_2=10^{-4}, a_3=10^{-2}$ for (c).
The behavior of very small amplitude, say,
$\log(C_{m})$ approximately  less than $-10 $  in panel (a),
mostly comes from numerical truncation errors, and
is therefore unimportant. }
\label{fig:mathmodel}
\end{figure}

%
The time-evolution of the amplitudes 
$\log(C_{m})\equiv \log_{10}(\sqrt{a_{m}^{2}+b_{m}^{2}})$ 
for some Fourier modes  $(m=1,\cdots , 6)$
is shown in Figure~\ref{fig:mathmodel}.
Figure~\ref{fig:mathmodel}(a) shows the results 
for $\lambda /\nu \approx 6.7$
($\lambda =1$, $\nu=0.15$). 
The linearized perturbation theory predicts no growing modes for this system.
The initial amplitudes are chosen as
$ a_2=a_3=10^{-2}$ and the others are zero.
Higher $m$ modes are always induced by the initial seeds,
$ a_2$ or $a_3$, but they all decay with time and their amplitudes
are small compared with the $ m=2$ and $m=3$ modes. That is, 
the $ m=2$ mode is the largest of the even modes and
the $ m=3$ mode is the largest of the odd modes.
%

Figure~\ref{fig:mathmodel}(b) shows the results for a less viscous model with
$\lambda /\nu =20$ ($\lambda=1,  \nu=0.05$).
There is one unstable mode, the $m=2$ mode,
predicted by linearized perturbation theory.
The initial conditions are the same as those 
in Figure~\ref{fig:mathmodel}(a).
The $m=2$ mode grows exponentially until $ t \sim 15$, where
the amplitude of the $m=2$ mode reaches the nonlinear regime:
$10^{-2} \times \exp(0.3 \times 15) \sim 1$.
The growth rate until $ t \sim 15$
agrees with $ \Gamma _2 =0.3$
derived from the linearized theory.
All other even modes,
originating from the bilinear coupling term
$ u \partial _x u  $, also grow.
The $m=6 $ mode is produced from the coupling between $m=2 $
and $m=4$ and also from the quadric coupling of $m=3 $.
Therefore, the amplitude of the $m=6$ mode is not always
smaller than that of $m=4$.
The growth of all even modes is slightly
suppressed after the turning time $t \sim 15$.
The turning time is also important for the odd modes.
The odd modes decay for $t <15$, but grow after that.
Therefore, the nonlinearity of the amplitude of the $m=2$ mode
cannot be ignored even for the odd modes.
The turning time corresponds to shock formation
as will be discussed later.
   In order to examine the effect of the nonlinearity of the
$m=2$ mode on the growth of all other modes,
we set the initial amplitude of $ a_2$ to $10^{-4}$.
The other initial conditions were kept the same as
those used to produce Figure~\ref{fig:mathmodel}(b).
The time-evolution shown in Figure ~\ref{fig:mathmodel}(c) has the same
 general features as Figure~\ref{fig:mathmodel}(b),
but the turning time, when the odd modes switch from decay to growth,
is shifted to $t \sim 30$.
This is because of the small initial amplitude of the $m=2$ mode:
$ 10^{-4} \times \exp(0.3 \times 30) \sim 1$.
For eqs.(\ref{eqn:Burgers}) and (\ref{eqn:Burgersforce})
all odd modes are always zero,
if they are exactly zero initially.
When there is at least one odd mode with a finite
amplitude, the nonlinearity of the $m=2$ mode enhances
all odd modes.

The nonlinear evolution for $\lambda /\nu =100$
($\lambda=1,  \nu=0.01$) is shown in Figure~\ref{fig:mathmodel}(d).
The initial conditions are the same as those used in 
Figure~\ref{fig:mathmodel}(b).
In this model the $m=2$ and $m=3$ modes are unstable with
growth rates of $\Gamma_2=0.46$ and
$\Gamma_3=0.053$ from the linearized theory.
Overall the features are the same as in Figure~\ref{fig:mathmodel}(b),
except for the timescale. The turning time due to the $m=2$ mode
is shorter in this model: $t \sim 10$
since $ 10^{-2} \times \exp(0.46 \times 10) \sim 1$.
The $m=3$ mode is initially unstable but
does not grow significantly during the early phase $ t <10$.
The typical growth timescale is very long,
$ \Gamma_3 ^{-1} \sim 20$, so that
mode coupling becomes much more important at early times.
However, the unstable $m=3$ mode maintains the amplitudes
of other odd modes at higher levels
through mode coupling before the turning time $t \sim 10$.

\subsection{Comparison with 3D Simulations}
 By comparing the mathematical model with 3D numerical 
results, the following features become clear.
The odd modes grow only after nonlinear saturation of
unstable mode. 
This fact can easily be seen in Figure~\ref{fig:diagnostic2}, 
 in which the initial amplitude of odd modes is $ 10^{-3}$ 
 and the nonlinearity becomes
 important in short timescale.
We further discuss the growth feature in
the models with small initial amplitudes.
In the mathematical model, 
the odd mode growth starts approximately from the time
$t_s + 2 \tau$, where $t_s $ and $ \tau$ are 
the saturation time and growth time of
the unstable mode. 
The starting point of the odd mode growth
is not easily determined in the actual calculations,
since the growth curve is not so sharp.
The similar relation is however realized.
The amplitudes of odd modes exceed, say, 
$\sim 5 \times 10^{-3}$ at the time  
$\sim t_s + 5 \tau$ in the models I-III 
in Figure~\ref{fig:diagnostic1}.
This property is almost independent of 
initial perturbations with $ \delta < 10^{-3}$ and unstable models.
The start time $t_g$ of odd mode growth is 
roughly given by  $t_g =t_s + \alpha \tau $,
where $ \alpha ={\mathcal O}(1) $.
The number $\alpha$ depends on 
dynamical degrees of freedom of the system.
The growth of the amplitude $a_m$ simply depends on the
bilinear coupling $\sum a_k a_{m-k}$ or $\sum a_k b_{m-k}$
in the mathematical model.
  There is a similar bilinear coupling in 3D simulation, but coupling
  is more complicated.  The azimuthal Fourier component of the density
  couples with three components of velocity, which couple with 
  those of energy, gravity and density.  Moreover, they are 
  functions of $x,y,z$.
The dynamical degrees are so large in 
the 3D system,
that the interval  $(t_g -t_s)/\tau  $ becomes longer.
The relation between the nonlinear saturation time and
growth time of odd modes qualitatively holds in both systems,
despite of the increased dynamical degrees of freedom.
  In the mathematical model, it is possible to
eliminate the initial perturbations of odd modes. 
In such a clean case, the odd modes can not appear.
In the actual numerical simulations, some random noises, whose 
amplitudes are expected as 
$ \sim 10^{-5} $-$10^{-4}$,  should be involved.
The noise reduction would be possible 
in future simulation on the much large scale computer, 
but is not necessary in the realistic applications
since such an ideal initial condition is rare.
Thus odd modes would appear in general.

\subsection{ Evolution of Shape}
\begin{figure}
\centering
 \includegraphics[width=7cm]{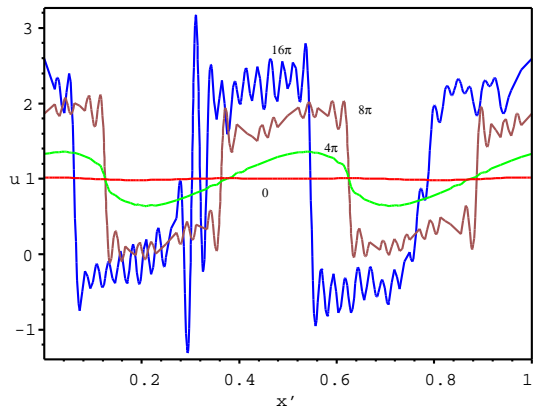}

 \caption{
Snapshots of the function $u(t,x)$  are shown
as a function of $x' = x/2\pi$ for
$ t=0, 4 \pi, 8\pi, 16\pi$.
The attached labels denote the time $t$.  }
\label{fig:velocity}
\end{figure}

 The similarity can be seen in the time evolution 
of the Fourier components both in mathematical 
and 3D numerical models as shown in previous subsection.
The time evolution of the shape $u(t,x)$ is shown 
in Figure~\ref{fig:velocity}.
The parameters and initial conditions are
the same as used to produce 
Figure~\ref{fig:mathmodel}(b). 
The snapshots are given for
times $t= 0, 4\pi, 8\pi$ and $16 \pi$.
The choice comes from removing the propagation effect,
because the initial velocity is $u \approx 1$.
The $m=2$ mode initially grows and the shape is enhanced
before the turning time  $t \sim 15 $.
The curve at $t=4 \pi $ clearly shows symmetric features
due to the $m=2$ mode.
That is, the shape is the symmetry under translations 
$ x \to x+ \pi$,
a ``$\pi$-symmetry''.
The nonlinearity causes a shock as in the original
Burgers' equation. After shock formation,
the Gibbs phenomenon associated with Fourier series
is seen at $t=8 \pi, 16 \pi $.
The overshoot is a numerical artifact and
such behavior always appears
when  a function having a sharp discontinuity
is expressed as a Fourier series\cite{AF01}.
Neglecting the Gibbs phenomenon,
the symmetry due to the $m=2$ mode can still be seen in the shape
at $t=8 \pi$, whereas it is partially broken at
$t=16 \pi $.
The time $t=16 \pi $ in the mathematical model
is much longer than that of non-linear saturation
and that of growth of odd modes.
Therefore, there is no counterpart in 3D
numerical simulations in Figures 1 and 2.
The mathematical model suggests that 
a ``$\pi$-symmetry''
(i.e., symmetry under a $180^{\circ}$ rotation around
the $z$-axis) in the shape
is broken in a longer timescale.
%

\section{Discussion}
\label{sec:Discussion}

  We have considered the nonlinear evolution
of the bar-type instability in a differentially rotating star
with significant rotational energy.
The previous numerical results 
of general relativistic simulation\cite{BPMR07}
suggest that the growth is likely
to come from mode coupling.
In order to obtain further evidence of mode coupling,
we have developed a simulation of three-dimensional hydrodynamics in 
Newtonian gravity and a simple mathematical model.
Our mathematically simplified model provides a concrete 
example showing the importance of mode coupling.
The amplitudes of odd modes increase without 
unstable odd modes being present in the axially symmetric state; 
instead, they are enhanced by the bar instability with $m=2$.
We also confirmed that this physical picture
is consistent with the results from a
three-dimensional hydrodynamics simulation.
Generally, the odd modes grow only after the bar instability
reaches the nonlinear regime.
The timescales of the mode coupling and 
the growth of unstable modes may depend on the rotation law and
the strength of the initial instabilities.
It is very rare
that the initial perturbations 
in the hydrodynamics simulation should consist of
purely  even or odd modes only.
Therefore, the  unstable bar mode enhances 
the amplitudes of the all other modes at late times, no matter whether they 
are even or odd.

A similar mode coupling can be seen in numerical simulations
for the one-armed spiral instability\cite{OT06}
and the elliptical instability\cite{OTM07}
of rotating stars in Newtonian gravity. 
  The initial models and the growth mechanism are different, but the
  turbulent-like behavior appears in diagnostics of
  the azimuthal Fourier components 
  at late times of nonlinear growth\cite{OTM07,CQF07}. 
  The behavior is also important for the
  nonlinear saturation of the unstable mode.  Further study is
  necessary to explore the origin of the similarity seen in the
  development of different unstable modes.
It is reasonable to assume that the nonlinearity in hydrodynamics 
is the source of this similarity.
What is the effect of general relativity?
A number of nonlinearities occur in general relativity which may
affect the growth of the unstable bar mode.
Although the time-evolution in full relativistic calculations
is very similar to that in Newtonian gravity\cite{BPMR07},  
it will be very interesting to explore further 
whether or not a full relativistic simulation 
produces a nonlinearity different from the
one presented by the simple model.

\acknowledgments
This work was supported in part 
by a Grant-in-Aid for Scientific Research (No.16540256) from 
the Japanese Ministry of Education, Culture, Sports,
Science and Technology.
MS thanks 
Luciano
Rezzolla and Shin Yoshida for discussion.  MS also thanks Misao Sasaki for
his kind hospitality at the Yukawa Institute for Theoretical Physics,
where part of this work was done. This work was supported in part by
the STFC rolling grant (No.~PP/E001025/1)
at the University of Southampton, 
by the Special Fund for Research program in
Rikkyo University, and 
by the
Grant-in-Aid for the 21st Century Center of Excellence in Physics at
Kyoto University.
Numerical computations were performed on the
myrinet nodes of Iridis compute cluster in the University of
Southampton, on the cluster in the Institute of Theoretical
Physics, Rikkyo University, and on the Cray XT4 cluster in the Center
for Computational Astrophysics, National Astronomical Observatory of
Japan.


\end{document}